\documentclass[12pt]{article}

\usepackage{ucs}
\usepackage{booktabs,longtable,tabu}
\usepackage[flushleft]{threeparttable}
\usepackage{threeparttablex}
\usepackage{dcolumn}
\usepackage[sc]{caption}
\usepackage{afterpage}
\usepackage{subcaption}
\usepackage{pgfplots}
\usepackage{latexsym,amsfonts,amssymb,amsmath,amsthm,mathrsfs,mathtools,dsfont}
\usepackage{bm}
\usepackage{enumerate}
\usepackage[english]{babel}
\usepackage[T1]{fontenc}
\usepackage[utf8]{inputenc}
\usepackage{graphicx}
\usepackage{epstopdf}

\usepackage{newpxtext}
\usepackage[noamssymbols,nosymbolsc]{newpxmath}

\usepackage{pdflscape}
\usepackage{pdfpages}
\usepackage{setspace}
\usepackage{verbatim}
\usepackage{grffile}
\usepackage{bigfoot}
\usepackage[round]{natbib}
\usepackage{footmisc}
\usepackage[left=1.2in,right=1.2in,top=1in,bottom=1in]{geometry}
\usepackage{multirow}
\usepackage{authblk}
\usepackage{tikz}
\usetikzlibrary{arrows}

\usepackage{color}
\usepackage[pdftex]{hyperref} 
\definecolor{darkblue}{rgb}{0.0,0.0,0.5}
\hypersetup{colorlinks,breaklinks,
            linkcolor=darkblue,urlcolor=darkblue,
            anchorcolor=darkblue,citecolor=darkblue}

\usepackage{textcomp}

\theoremstyle{definition}

\usepackage[nolists,tablesonly]{endfloat}  
\DeclareDelayedFloatFlavor{ThreePartTable}{table}
\DeclareDelayedFloatFlavor{landscape}{table}

\title{A Memo on the Proof-of-Stake Mechanism\thanks{Special thanks to Eric Budish, whose paper laid the foundation for this memo.}} 
\author{George Gui  $\quad$ Ali Horta\c{c}su $\quad$ José Tudón}
\date{First version: May 12, 2018 \qquad This version: \today}

\begin{document}

\maketitle
\begin{abstract}

We analyze the economic incentives generated by the proof-of-stake mechanism discussed in the Ethereum Casper upgrade proposal. Compared with proof-of-work, proof-of-stake has a different cost structure for attackers. In \cite{budish_economic_nodate}, three equations characterize the limits of Bitcoin, which has a proof-of-work mechanism. We investigate their counterparts and evaluate the risk of double-spending attack and sabotage attack. We argue that PoS is safer than PoW agaisnt double-spending attack because of the tractability of attackers, which implies a large ``stock'' cost for the attacker. Compared to a PoW system whose mining equipments are repurposable, PoS is also safer against a sabotage attack. 

\end{abstract}


\section{Introduction}

Under proof-of-stake (PoS), networks achieve distributed consensus on a blockchain by randomly choosing the creator of the next block as function of their stake. The stake is an amount of the cryptocurrency which is temporarily frozen. If the creator misbehaves, his stake is burned. Since an attacker faces a large ``stock'' cost of misbehaving, PoS is safer than proof-of-work (PoW), which generates a smaller ``flow'' cost for the attacker.

We assume that the reader has a basic knowledge of Bitcoin and its PoW mechanism. Detailed descriptions can be found in \cite{athey_bitcoin_2016}, \cite{huberman_monopoly_2017}, and \cite{budish_economic_nodate}. We (1) give a high-level description of the PoS mechanism discussed in the Casper proposal; (2) consider potential attacks; (3) discuss the connection to \cite{budish_economic_nodate}; (4) describe some possible solutions to the attacks and their implications; and (5) analyze equilibria. There are different flavors of PoS. In this memo, we mainly focus on the Casper proposal discussed in \cite{buterin_casper_2017}. 


\section{Proof-of-Stake}
In  PoS, validators vote on the authentic transactions based on their stake, an amount of Ethereum that they deposit into an account, which is frozen for a certain period of time. If they are identified as an attacker, however, they lose their deposit. In contrast, under PoW, validators are randomly selected based on the computation power they use. This process costs electricity that prevents attackers to change transaction records. The criticism of this electricity consumption leads to the invention of PoS. In this paper, we will not discuss the environmental advantage of PoS. Instead, we will focus on the tractability of attackers and the scarcity of stakes that leads to a different cost structure for an outside attacker. 


\subsection{Validation procedure}
For the rest of the paper, all the validators will be weighted by their deposits. When we say "$\frac{2}{3}$ of validators", we refer to the set of validators whose deposits are equal to $\frac{2}{3}$ of the total deposits. The proof-of-stake in \cite{buterin_casper_2017} can be described as follows.

\begin{enumerate}
\item Each block is 10-20 seconds and every 100th block is a checkpoint.

\item At each checkpoint, Ethereum owners can choose to stake Ethereums to become validators and earn rewards. Validators can choose to exit, but cannot withdraw their deposits until 4 months later.

\item Because of network latency, conflicting blocks may occur. 
 Validators are incentivized to coordinate on which checkpoints should be added to the chain as the canonical history. This coordination is done by broadcasting their intentions to vote for multiple rounds subjected to certain rules.

\item A checkpoint that is based on a valid history is \textbf{finalized} if $\frac{2}{3}$ of the validators\footnote{The number  
 is chosen based on Byzantine fault tolerant consensus theory. } vote for the checkpoint. 
 
\item When a checkpoint is finalized, merchants are expected to trust that the transactions will not be reverted unless $\frac{1}{3}$ of validators maliciously double vote. By double voting, attackers can trick a merchant to accept the Ethereums that they have already spent and finalized elsewhere. Honest validators are incentivized to report such behaviors and burn the Ethereum deposits of the attackers. 

\item When two finalized checkpoints conflict, because more than $\frac{1}{3}$ of validators misbehave, offline coordination\footnote{For example, organizing a vote on Twitter or some other trustworthy websites.} is expected to resolve the issues. Only one version will be selected in the end.
\end{enumerate}

\section{Potential Attacks}


\begin{description}
\item[Double-Spending (Finality Reversion)]: The process is similar to blockchain. 
An attacker (1) acquires $\frac{2}{3}$ of stakes; (2) sends his Ethereum to a merchant in exchange for goods or assets by voting to finalize a checkpoint that includes the transactions; 3) finalizes a conflicting checkpoint that sends the same Ethereum to another merchant (double-spends); (4) gets caught and his stakes are burned as honest validators are incentivezed to report such misbehavior. 

 Note $\frac{1}{3} + 2\epsilon$ is enough for the attack if the network latency is high. Suppose the attacker owns $\frac{1}{3} + 2\epsilon$ of stakes, validators in Europe owns $\frac{1}{3} - \epsilon$ of stakes, and validators in Africa own the remaning $\frac{1}{3}  - \epsilon$  of stakes. The attacker can vote for the transaction in Europe with the European validators, and then vote for the conflicting transaction in Africa with the African validators using the same Ethereums. Then both transactions will be finalized because they have $\frac{2}{3} + \epsilon$ votes. However, Merchants in both Africa and Europe may have only observed one finalized checkpoint because of the network latency. Both of them will approve the transaction, but only one of them will get paid eventually. 

\item[Sabotage (Going offline)]: An attacker owning $\frac{1}{3} + \epsilon$ of the stakes can refuse to vote, which prevents checkpoints of being finalized, which brings transactions to a halt. Ethereums users are expected to coordinate outside of the network to censor the malicious validators, but it is an open question that remains unresolved. We will mentione some possible methods in the Appendix \ref{append:coordination}.\footnote{\cite{buterin_casper_2017} suggested that one way is to gradually down-weight validators who don't vote. But there is a likelihood that even if more than $\frac{2}{3}$ of validators are honest, sometimes there are conflicting checkpoints because honest validators accidentally go offline.} 
\end{description}

Both of the attacks can induce conflicting finalized checkpoints and will require offline coordination by honest users. We will discuss the potential difficulty and cost of the coordination in Appendix  \ref{append:coordination}. 

\section{Connection to Budish (2018)}\label{sec:budish}

\subsection{Competition Among Miners}
Let $P_{block}$ be the reward for mining an Ethereum block, $N$ the total stakes deposited into the frozen account, and $c$ the opportunity cost of not being able to spend the frozen stakes. Then, we can calculate the equilibrium amount of stakes in the system $N^*$ with
\begin{equation}\label{eqn:reward}
N^*c = P_{block} 
\end{equation}  

With PoW, the cost mostly involves renting the mining machines and the variable cost of the electricity. This equation is analogous to the first equation in \cite{budish_economic_nodate},  where, analogously, $N^*$ is the computational power devoted to mining, and $c$ is the per-block unit cost of the computational power. 
 
%

\subsection{Incentive Compatibility} 

Both double-spending and sabotage attacks will lead to losing all the stakes.\footnote{It is also possible for the attackers to bribe other validators, The bribing techniques and the corresponding cost would be an interesting topic to study. But in this paper we mainly focus on the case when the validators need to own the Ethereum. } The equation should approximately be
\begin{equation}\label{eqn:incentive}
N^{attack} C > V_{attack}
\end{equation}

where $N^{attack}$ is the number of Ethereums needed to complete the attack; $V_{attack}$ is the value of the attack;  $C$ is the cost of the Ethereum. \footnote{The Ethereum cost $C$ is not strictly linear in $N^{attack}$ because of the limited supply of Ethereum. When the existing stakes are already more than $\frac{2}{3}$ of total Ethereum supply,  buying sufficient Ethereum for the attack is impossible. However, for simplicity, we assume away this appreciation.} For an outsider attacker, $N^{attack}$ is at least $50\%$ of the total existing stakes, so he will own at least $\frac{1}{3}$ when he deposits his additional $\frac{N^*}{2}$ Ethereums.

The stakes in PoS is analagous to the chips used to mine the reward in a PoW network. As discussed in \cite{budish_economic_nodate}, they fit into the case where the flow cost approach is not appropriate because "the most efficient chips are specialized, there are neither reasonably efficient repurposable chips nor older generation specialized chips". 

%
%

\subsection{Ethereum Cost as a Function of Reward} 
If we view profit generated by depositing the Ethereum as dividend or interest, then the Ethereum price should equal to the sum of discounted profit over a infinite horizon. Thus we can draw the connection between the opportunity cost $c$ and the Ethereum price $C$. 
 Let $\beta$ be the discount rate, then: 
\begin{equation}\label{eqn:dividend}
C = c\sum_{t = 1}(1 + \beta + \beta^2 + \beta^3 + ...) = \frac{c}{1-\beta}.
\end{equation}

Both double-spending and sabotage attacks require at least half of the existing stakes, so $N^{attack} = \frac{1}{2}N^*$. Attackers do not have the incentives to attack when the value of the attack is smaller than the cost of buying and losing the stakes: 
\begin{equation}\label{eqn:conversion}
\begin{split}
V_{attack} & < N^{attack} C = \frac{1}{2}N^* C\\
 &=  \frac{1}{2(1-\beta)}N^*c
\end{split}
\end{equation}

Note the qualitative similarities of both sides of equation \eqref{eqn:conversion}: both sides have \emph{stock} costs. That is, a one-shot gain of an attack must be compared with a perpetual cost. Compare with the case of PoW, where costs are a flow \citep{budish_economic_nodate}. Equation \eqref{eqn:conversion} implies a higher degree of security for PoS, as stock costs are higher than flow costs.

Appendix Section \ref{sec:miners_too_low} argues the long-run equilibrium number of miners is low, yielding less security. Appendix Section \ref{sec:interest_rate} analyzes the equilibrium interest rate implied by equation \eqref{eqn:dividend}.



\subsection{Cost Comparison} 

For notational convenience, let   
\begin{equation}\label{eqn:alpha}
\alpha_{ether} = \frac{1}{2(1-\beta)}
\end{equation}

If we assume an annualized discounting factor of $0.98$, and $\beta$ is the discounting factor of a 10-minute period, then $\beta$ will be approximately $0.9999996$. 
The correpsonding $\alpha_{ether}$ is around $10^6$.

Combing equation (\ref{eqn:reward}), (\ref{eqn:conversion}) and (\ref{eqn:alpha}), we can get 

\begin{equation}
P_{block} > \frac{V_{attack}}{\alpha_{ether}}
\end{equation}

This equation has a similar format as the one discussed in \citep{budish_economic_nodate}, except that the composition and magnitude of $\alpha_{ether}$ are different. 
In both of PoW and PoS, $\alpha$ can be understood as the cost for attackers to misbehave relative to the cost for honest validators to earn block reward. 

In PoS, the honest validators only pay the opportunity cost of depositing the stakes, but attackers, both double-spenders and saboteurs, lose their stakes. 

In PoW, the flow cost of a double spending attack is not qualitatively different from the flow cost of maintenance for honest validators, as electricity is the only  cost. An example number used by \citep{budish_economic_nodate} for the Bitcoin coefficient $\alpha_{\beta}$ is $3.35$, which will be several magnitude smaller than $\alpha_{ether}$. On the other hand, the cost of a sabotage attack depends on how repurposable the mining chips are. If chips are repurpusable, then their preserve value after the network is sabotaged. Thus, the cost of sabotage is mainly a flow cost from electricity, and is not qualitatively different from that of honest miners. However, if chips are not repurposable, then the cost of sabotage includes the cost of the mining chips as well. Thus, the cost becomes a stock cost.

If Ethereum prices are linear, sabotaging under PoS is cheaper: if we regard stakes as mining chips, an additional 100\% of the current mining power is needed for a PoW sabotage attack, but only 50\% of the current stakes are required for the PoS attack.\footnote{This description holds true for the  algorithms currently used by Bitcoin and Ethereum.}

%

However, if we consider the possibility of offline coordination that can possibly resolve the attack and identify the malicious behavior, a PoS network becomes more expensive to attack. In PoS, the network can analyze the behavior of validators to identify the malicious ones. In PoW,  tracking down the malicious mining machines and destroying them is difficult. Therefore, even unsuccessful attacks incur in a large stock cost under PoS.

\bibliographystyle{chicago}
\bibliography{pos_bib}

\appendix

\section{Coordination and Possible Solutions}\label{append:coordination}
If double spending or sabotage attack takes place in PoS, coordinations outside of the network are needed to resolve the conflict.\footnote{In Bitcoin offline coordinations are not needed because when there exists two blockchain, the longest one will be chosen.} Coordinations are not rare at the current stage of the cryptocurrency development. All kinds of forks and upgrades for both Bitcoin and Ethereum require offline coordinations, usually on Twitter or other websites. While attackers are unlikely to have incentives to double spend based on discussion in Section \ref{sec:budish}, they still have the incentives to sabotage the system if the future market is large. There are possible ways to deter the sabotage attacks, with their own limitations. 

\subsection{Coordination on eliminating malicious validators}
When an attacker owns $\frac{1}{3}$ of the deposit, he can intentionally pretend to be offline and not voting for any block, resulting in a sabotage attack. Because in PoS it is possible to track an validator, it is possible to analyze likely malicious behaviors and censor them from the network. Other honest validators, include exchanges, can eliminate or censor these malicious validators by creating a soft fork of the Ethereum blockchain that excludes these validators.

This potential punishment can deter a sabotage in multiple ways. For example, if the attackers target the future market, the future contract will then be based on the price of this forked chain. In this forked chain, because the valid supply of Ethereum decreases instead, the price of the Ethereum can possibly increase. Thus, the attacker will lose both from both the Ethereum stake and the future contract. Conditional on successful coordination, the potential loss from future contracts is linearly proportional to the value of the attack. The attackers will take into account the risk tolerance and the probability of successful coordination when initiating the attack.

Nevertheless, when a validator goes offline for a period $t$, it is unclear if he is malicious or his node experiences some network failure. There also exists other malicious behavior that will not get punished directly according to the rules. The detection algorithm of such ambiguously malicious event can mistakingly penalize honest players, lowering the expected payoff and increasing the risk of an honest validator.

\subsection{Middlemen Cartel}
Because stakes are scarce in the PoS mechanism, if the remaning liquid Ethereum is below $50\%$ of the existing stakes, it becomes imposible for an outsider to attack. When the optimal level of deposit is higher than $\frac{2}{3}$ of the total Ethereum, and a group of firms that run Ethereum-related business and have the incentives to keep the system safe, then one strategy that they can pursue is to own all of the deposits and promise consumers to behave honestly. For example, large exchanges whose businesses rely on cryptocurrency have the incentive to become coordinators when the system is being attacked. The settlement prices of the current Bitcoin future contracts are all based on the exchanges. Because all of the information is public, consumers and coordinating firms can check if everyone keeps their promises. Because the identity is revealed, it is no longer a one-shot game and there can also be legal consequences if they break the promise. Of course, it is unclear how different this centralized middleman is from a traditional middleman. This cartel can censor transactions that are unfavorable to them or limit the number of transactions to increase profit. One key difference is that all of these behaviors are transparent and easier for the public to supervise.  If this network has lower cost, traditional middlemen can run such network in its backend. Under this setting, Ethereum will in some way look like another cryptocurrency, Ripple, which is run by a coalition of enterprises. 
 
\section{Competition with Traditional Middlemen: an Upper Bound for Transaction Fees} \label{sec:miners_too_low}

If coordinations outside of the network are arbitrary and can even include governemnt and large firms who have interest in the network, it is unclear how much advantage the network has left compared to the traditional payment system.

Currently, miners in either systems are rewarded for adding to the blockchain with a block-reward, which consists of two parts: a number of newly minted coins, and the block transaction fees. In the long-run, the number of newly minted coins will approach zero.\footnote{Bitcoin minting will stop around the year 2140.} 
Thus, the reward will be mainly composed by transaction fees, which is linear to the security of the system. However, transaction fees are disciplined by the outside option, which consists of transferring money through traditional payment systems. Since services such as Visa and Mastercard are cheap to run, Bitcoin and Ethereum will either charge low transaction fees, or be used in scenarios in which traditional transaction methods are expensive. However, for both algorithms, the long-run equilibrium number of miners will be low because transaction fees will be low. Thus, both PoS and PoW will become less safe in the long-run.

For Ethereum, its smart contract can provide trust for supporting gambling or risk-hedging activities that now still charge relatively high transaction fee. As sports gambling has become legal in the U.S., Ethereum can still possibly sustain large transaction fee to maintain the security of the system. This smart contract feature is orgothonal to the PoS mechanism but will be an interesting topic to study in the future. 

\section{Opportunity Cost of a Deposited Ethereum}\label{sec:interest_rate}

This exploratory section is further investigates the opportunity cost of deposited Ethereums. 

Suppose that, in equilibrium, the total number of Ethereum does not change. 
\begin{equation}
N_{total} = N_{deposit}+ N_{liquid}
\end{equation}

The value of a liquid Ethereum partly lies in the ability to sell them when the demand surges. 

As in \cite{athey_bitcoin_2016}, we can use the notion of velocity to establish the connection between supply and the exchange rate. 
\begin{equation*}
\text{Exchange rate} = \frac{\text{Transaction Volume}}{\text{Velocity} \times \text{Supply of Liquid Ethereums}}
\end{equation*}

In simpler notations
\begin{equation}\label{eqn:velocity}
p = \frac{D}{v  \times N_{liquid}}
\end{equation}

where $D$ is the dollar transaction value, $p$ is the exchange rate, or the price of Ethereum, and $v$ is the velocity term that describes how easy it is to use this currency.

We assume the following: (1) The velocity is approximately constant, because the opportunity to use them and the corresponding technology has become stable; and (2) demand shocks $\Delta D$ are i.i.d.

Then the expected payoff of a unit Ethereum based on the demand volatility will be
\begin{equation}\label{eqn:liquid_payoff}
\begin{split}
r_{liquid} &= P(\Delta p > 0) E[\Delta p|\Delta p > 0] \\
 & = P(\Delta D > 0) E[\frac{\Delta D}{v\times N_{liquid}}|\Delta D > 0] \\
 & = \frac{1}{N_{liquid}}P(\Delta D > 0) E[\frac{\Delta D}{v}|\Delta D > 0]\\
& = \frac{1}{N_{liquid}}P_{volatility}
\end{split} 
\end{equation}
where we let $P_{volatility} = P(\Delta D > 0) E[\frac{\Delta D}{v}|\Delta D > 0]$  denote the total expected payoff generated by demand volatility. In equation \eqref{eqn:dividend}, the discount factor $\beta$ can be interpreted as $r_{liquid}$.

This equation says that the expected payoff a liquid ethereum is inversely proportional to the total liquid supply, conditional on a demand shock distribution. 

Equation (\ref{eqn:reward}) $N^*c = P_{block}$ suggests that the unit profit of a deposited Ethereum is inversely proportional to the total number of deposit.

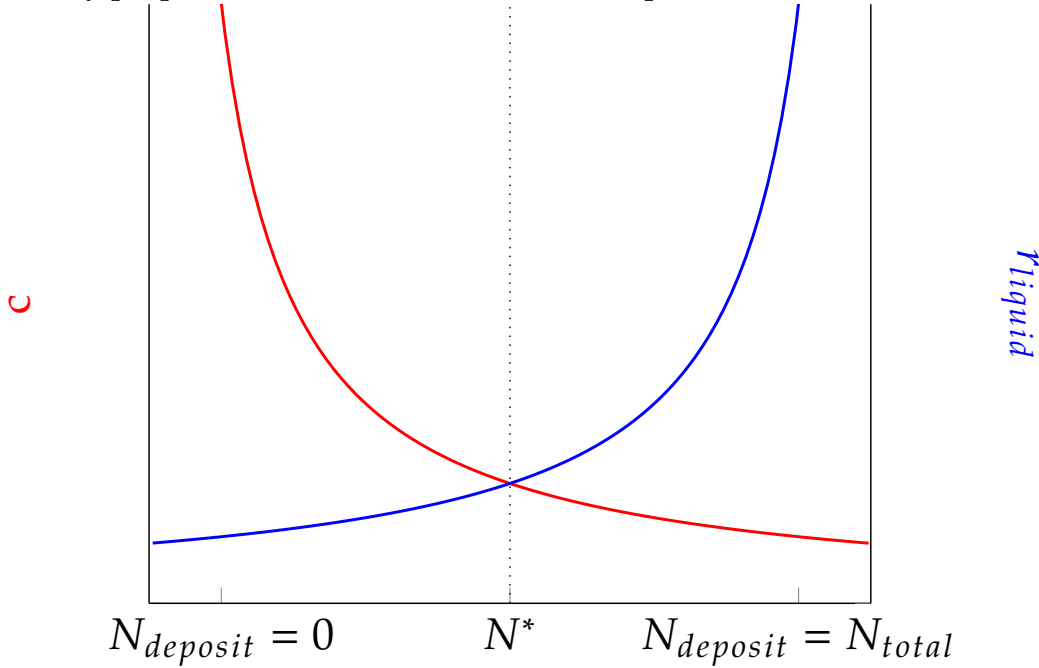
\begin{tikzpicture}[scale=1.4, transform shape]
\begin{axis}[ytick={0},  axis x line*=bottom, axis y line*=left, xmin=0, xmax = 2, ymin= 0, ymax = 2,
    samples=500,  
    xtick={0.2, 1, 1.8},
    xticklabels = {$N_{deposit} = 0$, $N^*$, $N_{deposit} = N_{total}$},
    ylabel = \textcolor{red}{c},
    yticklabels = {,,} , 
    legend style={draw = none, at={(0.5,-0.15)},
		anchor=north, legend columns=-1}
    ]
\addplot [color=red, thick][restrict x to domain=0:2] {0.4/x}; 
\addplot [color=blue, thick][restrict x to domain=0:2] {0.4/(2 - x)}; 
      \addplot +[mark=hello] [color=black, dotted] coordinates {(1, 0) (1, 2)};
      \legend{$c = \frac{P_{block}}{N_{deposit}}$, $r_{liquid} = \frac{P_{volatility}}{N_{liquid}}$}
\end{axis}
\pgfplotsset{every axis y label/.append style={rotate=180,yshift=9.5cm}}
\begin{axis}[
     xmin=0, xmax=2,
     ymin=0, ymax=2,
     hide x axis,
     axis y line*=right,
     yticklabels = {,,},
     ytick = {0} ,
     ylabel={\textcolor{blue}{$r_{liquid}$}}
 ]
\end{axis}

\end{tikzpicture}

Finally, the current level of deposit is determined by the demand distribution, the velocity of the currency, and the block reward, $P_{block}$.

\end{document}